# A Germanium-Vacancy Single Photon Source in Diamond

Takayuki Iwasaki,[1,2] Fumitaka Ishibashi,[3] Yoshiyuki Miyamoto,[2,4] Yuki Doi,[3] Satoshi Kobayashi,[3] Takehide Miyazaki,[2,4] Kosuke Tahara,[1] Kay D. Jahnke,[5] Lachlan J. Rogers,[5] Boris Naydenov,[5] Fedor Jelezko,[5] Satoshi Yamasaki,[2,6] Shinji Nagamachi,[7,8] Toshiro Inubushi,[8] Norikazu Mizuochi[2,3] and Mutsuko Hatano[1,2]

[1]Department of Physical Electronics, Tokyo Institute of Technology, Meguro, Tokyo 152-8552, Japan

[2]CREST, Japan Science and Technology Agency, Chiyoda, Tokyo

[3]Graduate School of Engineering Science, Osaka University, Toyonaka, Osaka 560-8531, Japan

[4]Nanosystem Research Institute, National Institute of Advanced Industrial Science and Technology, Tsukuba, Ibaraki 305-8568, Japan

[5]Institute for Quantum Optics and Center for Integrated Quantum Science and Technology (IQst), Ulm University, Albert-Einstein-Allee 11, Ulm D-89081, Germany

[6]Energy Technology Research Institute, National Institute of Advanced Industrial Science and Technology, Tsukuba, Ibaraki 305-8568, Japan

[7]Nagamachi Science Laboratory, Amagasaki, Hyogo 661-0976, Japan

[8]Shiga University of Medical Science, Otsu, Shiga 520-2192, Japan

Color centers in diamond are widely recognized as a promising solid state platform for quantum cryptography and quantum information processing. For these applications, single photon sources with a high intensity and reproducible fabrication methods are required. Here, we report a novel color center in diamond, composed of a germanium (Ge) and a vacancy (V) and named the GeV center, which has a sharp and strong photoluminescence band with a zero-phonon line at 602 nm at room temperature. We demonstrate this new color center works as a single photon source. Both ion implantation and chemical vapor deposition techniques enabled fabrication of GeV centers in diamond. A first-principles calculation revealed the atomic crystal structure and energy levels of the GeV center.

Single photon sources are a valuable resource for quantum cryptography[1] and quantum information processing[2], and impurity-related optical centers in diamond are promising candidates. Although many optically active structures have been found in diamond[3], only a limited number have been reported as a single photon source[4,5], such as Nitrogen-vacancy (NV)[6–8], Silicon-vacancy (SiV)[9,10], NE8[11], and Cr-related[12] centers. Of these, only the NV and SiV centers have been reproducibly fabricated[13]. To achieve superior optical properties and to more deeply understand the formation mechanism and characteristics of color centers in diamond, further exploration of novel color centers which can be reproducibly formed and have the single photon emission capability is required. In this study, we demonstrate that a germanium-related complex fabricated in diamond shows a sharp and strong luminescence band with a zero phonon line (ZPL) at around 602 nm, and has single photon emission capability at room temperature. Using first principle calculations, we found this color center to be composed of a Ge atom and a vacancy, namely GeV center, with the Ge atom relaxing to the bond-centered position giving $D_{3d}$ symmetry as in the SiV center. As well as production by ion implantation, we also demonstrate the formation of the GeV centers in diamond by chemical vapor deposition (CVD) and show that this leads to narrower line widths and smaller variation of the peak position. Theoretical calculation of the expected energy levels has revealed the reason for fluorescence energy difference from the SiV center.

The luminescent Ge-related structure was prepared by ion implantation of Ge ions into diamond and subsequent annealing. Figure 1a shows a PL spectrum from a Ge ion-implanted diamond at room temperature. A peak was found at 602.7 nm (~2.06 eV), accompanying a Raman scattering peak from the diamond substrate. We confirmed that the peak appeared over the whole substrate surface although the intensity varied (Fig. 1b) and that the peak intensity increased with increasing the Ge ion implantation dose (see Supplementary Fig. S1). This peak was therefore concluded to be the ZPL of the Ge-related structure in diamond. The full width at half maximum (FWHM) of the peak decreased at 10 K, where the ZPL splits to two components with an energy separation of 0.67 meV (inset in Fig. 1a). Other lines appear around the two peaks (for example at 601 nm), but more detailed study would be necessary to determine the fine structure. This luminescence band was only visible after the high-temperature treatment at 800 ºC and above. Ge ion implantation alone did not lead to the appearance of the peaks (see Supplementary Fig. S2). This fact indicates that the Ge forms a complex in diamond with a vacancy or vacancies diffusing during the high temperature annealing process, like other color centers related to vacancies[4].

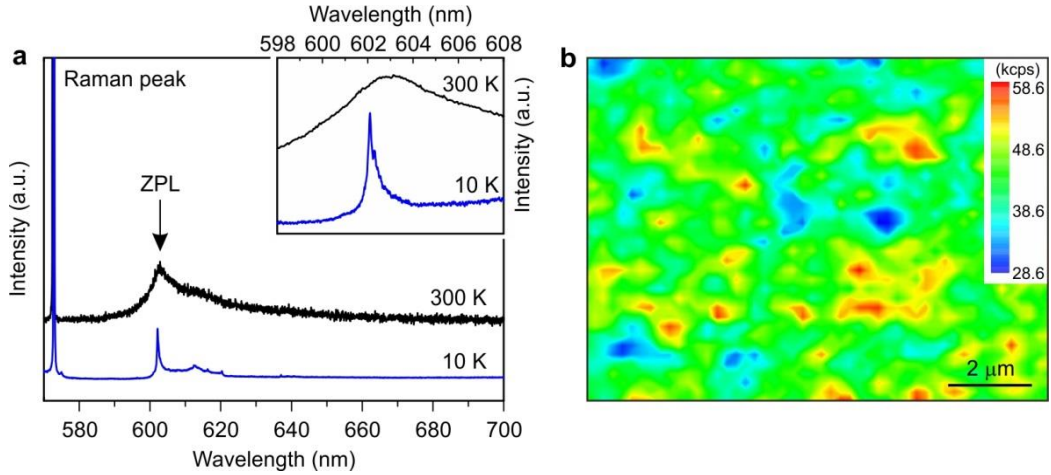

**Figure 1.** Luminescence characteristics of GeV color center in diamond formed by ion implantation. (a) PL spectra from a Ge ion implanted diamond at 300 K and 10 K. The inset shows ZPL at both the temperatures. (b) Intensity mapping of the ZPL between 595 and 608 nm at room temperature. The Ge ions were implanted to give a peak concentration of $1\times10^{19}$ cm$^{-3}$. The ion implantation conditions were determined by simulating SRIM[14]. The measurements were done by using a micro-Raman system at 300 K and a micro-PL system at 10 K.

Here, we demonstrate that the GeV color center works as a single photon emitter. Figure 2a, b show PL maps of a sample prepared by ion implantation with a much lower peak concentration ($1\times10^{14}$ cm$^{-3}$). A number of small fluorescent spots are observed in both the images. Two individual spots with a size of around 350 nm, marked in the white circles, show clear ZPLs from the GeV center, having peak positions of 601.6 and 602.4 nm for emitter 1 and 2, respectively (Fig. 2c). We measured the second-order autocorrelation function $g^2(\tau)$[4,8] for these spots with a band-pass filter around the ZPL from the GeV center (Fig. 2d). Sharp dips at a delay time τ of 0 ns for both the emitters indicate antibunching, and g²(0) below 0.5 is proof of single photon emission[15] from the GeV centers. The g²(τ) data were fitted with the equation for a three-level system[13], $g^2(\tau) = 1 - (1+\alpha)e^{\frac{-|\tau|}{\tau_1}} + \alpha e^{\frac{-|\tau|}{\tau_2}}$, where $\tau_1$, $\tau_2$, and α are the fitting parameters. We obtained a value of 1.4 – 5.5 ns for $\tau_1$, which provides an estimate of the excited state lifetime. This is much shorter than the NV centers (~12 ns)[6,16] and comparable to the SiV centers (1.1 – 2.4 ns)[9] in the bulk diamond.

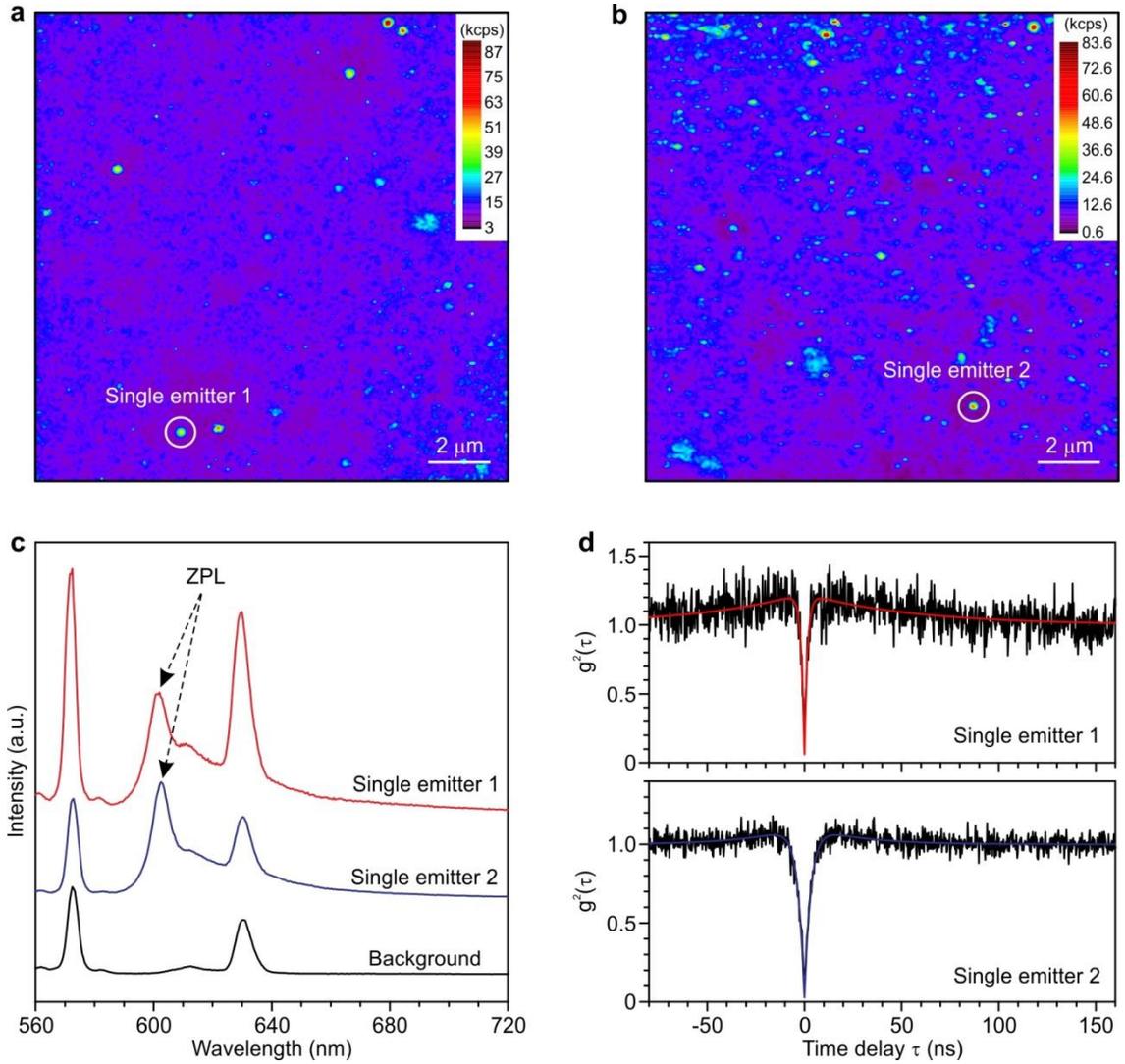

**Figure 2.** GeV single photon source. (a,b) Intensity mappings of GeV single emitters. (c) PL spectra

and (d) $g^2(\tau)$ function of the two GeV single centers, marked in the white circles in panels a and b. The background PL was collected at a position without a GeV center. The PL spectra were measured at an excitation laser power of 3 mW. Their intensities were normalized at ZPL. The $g^2(\tau)$ functions were measured at an excitation laser power of 1 mW. The solid lines in panel d denote the fitting. The sample was prepared by ion implantation at a dose of $3.5\times10^8$ cm$^{-2}$ and an ion energy of 150 keV, leading to a projected range of 57 nm from the surface. All the measurements were performed by using a confocal microscope system at room temperature. The confocal images and $g^2(\tau)$ functions were observed with a band-pass filter of 25 nm FWHM around 600 nm.

To analyze the behavior of the GeV single photon source in more detail, we measured the excitation power dependence of the $g^2(\tau)$ function (Fig. 3a). The antibunching was observed at the various laser powers, while increased excitation power led to the generation of maximum points over the unity. This bunching suggests that storage mechanisms are present in the GeV center, such as a shelving state or photo-ionization. The photon count rate is an important figure of merit for single photon sources. Saturation curves of these two emitters are shown in Fig. 3b. Fits of the form, $I = I_\infty \times P/(P + P_{sat})$[9], where $P_{sat}$ is the saturation power, reveal saturation intensities $I_\infty$ of 75 kcps (emitter 1) and 170 kcps (emitter 2). The value from the emitter 2 is slightly lower than 200 kcps reported for a SiV single emitter fabricated by CVD[17]. However, we measured the fluorescence only in the band of about 25 nm width around 600 nm due to a band-pass filter being required to avoid Raman signals around ZPL. It is expected that $I_\infty$ for GeV would be increased if the fluorescence could be detected across a wider spectral region. Therefore, the luminescence intensity of GeV is considered to be comparable to that of SiV. It should be noted that $I_\infty$ of NV$^-$ in the setup described in Ref. 17 is similar to that in the present measurement setup, making the comparison here between the GeV and SiV centers independent of the measurement system.

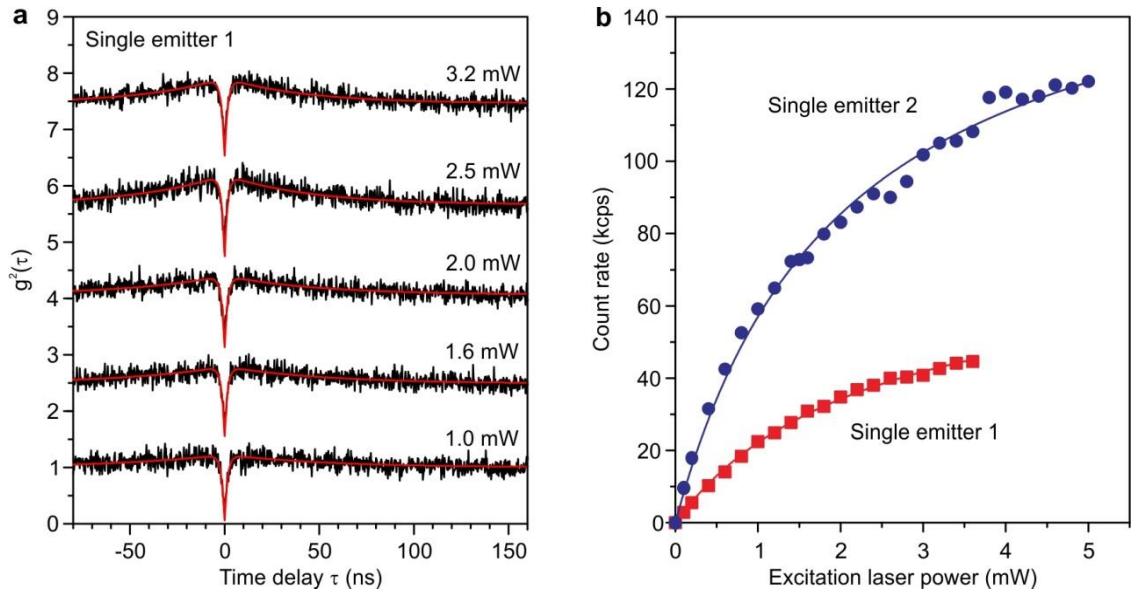

**Figure 3.** Analysis of GeV single photon source. (a) The $g^2(\tau)$ functions measured with different laser powers from 1.0 to 3.2 mW. The lines were shifted vertically for clarity. (b) Saturation characteristics of the two GeV single centers. The background was subtracted. The solid lines in panels a and b denote the fitting. All the measurements were performed by a confocal microscope system at room temperature with a band-pass filter of 25 nm FWHM around 600 nm.

CVD incorporation of color centers is an important technique to obtain high-quality fluorescence centers in diamond without implantation damages[10,17,18]. Here, we demonstrate the formation of GeV color centers in diamond by microwave plasma CVD (MPCVD). A diamond film was grown on a single-crystal diamond substrate by MPCVD with a Ge solid source. The Ge incorporation in the diamond film was confirmed by secondary ion mass spectrometry (SIMS) (see Supplementary Fig. S3). Figure 4a shows a PL spectrum from the CVD-incorporated GeV centers possessing a narrower line width of 4 – 5 nm than that (6 – 7 nm) of the GeV centers formed by ion implantation and annealing. Histograms of the ZPL position for GeV centers in the MPCVD and ion implantation samples are shown in Fig. 4b, and there is a slight blue-shift in the MPCVD sample. The CVD-prepared GeV centers also have a narrower inhomogeneous distribution ($\sigma$=0.05 nm) than those produced by ion implantation ($\sigma$=0.18 nm). These effects could arise from the lower defect density and lower strain in the sample prepared by MPCVD.

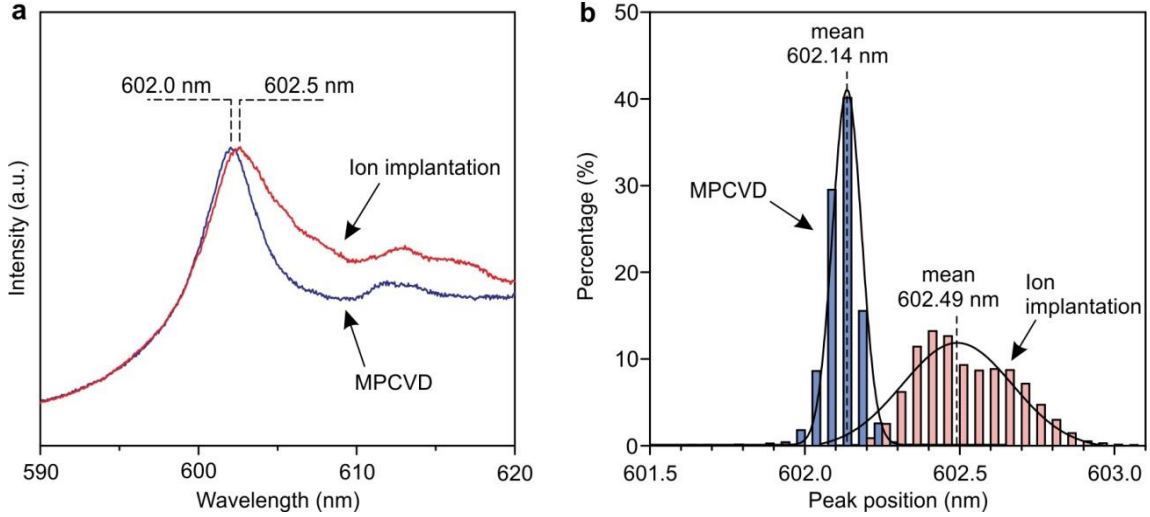

**Figure 4.** MPCVD-incorporated GeV color center ensemble. (a) PL spectrum from the MPCVD-incorporated GeV centers, compared with ones fabricated by ion implantation. (b) Histograms of the ZPL position of the GeV centers fabricated by MPCVD and ion implantation.

The crystal structure of the Ge-related color center was calculated from first-principles. First, we started from a structure assimilating to the NV center, i.e. one Ge atom at a substitutional site and one neighboring vacancy also at a carbon site. By performing structural relaxation, no potential barrier was found for the Ge atom on a trajectory from the substitutional site to an interstitial site between the lattice vacancies as shown in Fig. 5a, in agreement with previous theoretical study[19]. This geometry belongs to the symmetry group of $D_{3d}$, and it is the same configuration as the SiV center[20]. These atoms have similar electron configurations, and both prefer the interstitial position as they are substantially larger than the carbon atoms of the diamond lattice.

Energy levels of the SiV and GeV color centers were calculated within the density functional theory (DFT) using PBE functional to investigate the origin of the fluorescence wavelength difference (SiV has a ZPL at 738 nm[21]). Optical matrix elements were calculated to check optical selection rule among the computed levels. The calculation was done for the negatively charged SiV (SiV$^{-1}$)$^{22,23}$, neutral GeV (GeV$^0$), and negatively charged GeV (GeV$^{-1}$) centers. Figure 5b shows calculated energy levels of the SiV and GeV color centers in diamond. For both centers, the $e_u$ and $e_g$ levels are doubly-degenerate and the $e_g$ levels are partially occupied. The $e_u$ levels positioned in the valence band of diamond are at -0.67 eV for SiV$^{-1}$, -0.71 eV for GeV$^0$, and -0.48 eV for GeV$^{-1}$. Here, the energy is defined from the valence band maximum (VBM) of diamond. The $e_g$ levels are, however, in the band gap of diamond. The $e_g$ level of the SiV$^{-1}$ center is +0.74 eV above VBM, which agrees well with the previous calculation[24], while the GeV$^0$ and GeV$^{-1}$ centers have higher $e_g$ energies. Here, no significant difference in the energy between the $e_u$ and $e_g$ levels were seen for the GeV$^0$ and GeV$^{-1}$ centers. Although the charge state of the GeV center is not clear in this study, the higher $e_g$ levels should be the origin giving the higher fluorescence energy in the GeV center. Although the spin configuration is considered in the diagram according to Ref. 24, the current calculation was done under spin-unpolarized approximation[25].

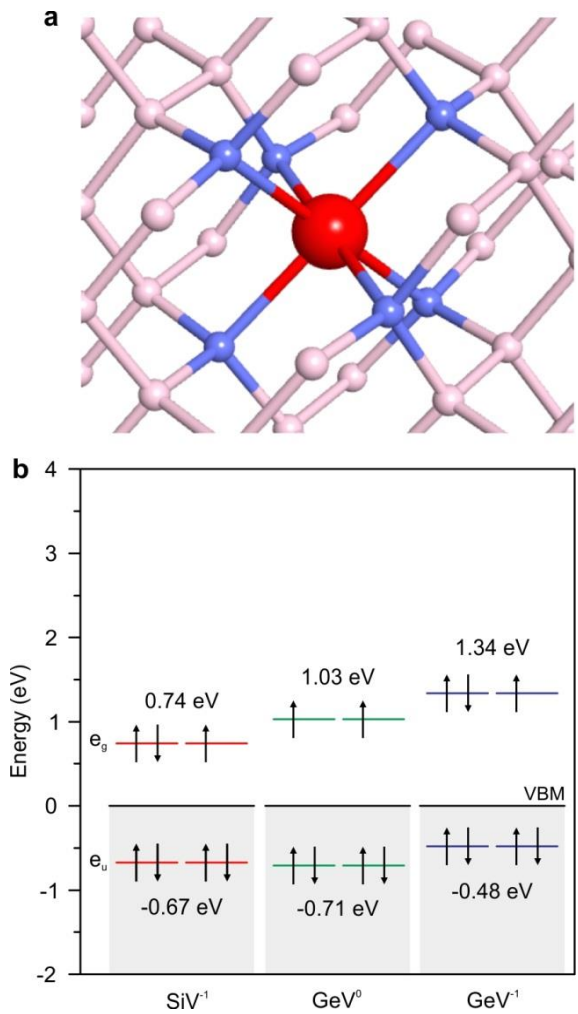

**Figure 5.** First principles calculation of GeV color center in diamond. (a) Crystal structure of the GeV color center in diamond. The red sphere denotes a Ge atom. The small blue and pink spheres are carbon atoms. The atomic structure shown is for the negatively charged state. (b) Energy levels of the SiV and GeV color centers in diamond. The energy was calculated with respect to VBM of diamond.

We have demonstrated that GeV centers can be reliably and reproducibly fabricated in diamond by ion implantation under various implantation conditions. Importantly, the capability of the single photon emission has been demonstrated. Additionally, it was confirmed that GeV centers can form by the incorporation of germanium during MPCVD growth, and these show less variation of the ZPL peak positions. These results establish the GeV center as a new single photon emitter that can easily be formed in diamond. Here, the GeV centers were fabricated in the bulk and thin film diamonds, but the morphology and size of diamond is in principle not limited. For example, the incorporation of the GeV centers in nanodiamonds should be possible, which is important for bio-labelling applications[26,27].

The large inhomogeneous distribution of the fluorescence wavelength of the GeV centers produced by ion implantation (shown in Fig. 4b) likely originates from the strain of the GeV complex structure in the diamond lattice. It is hard to completely remove residual defects, such as interstitial C atoms and vacancies, created during ion implantation by annealing. The remaining defects cause displacement and distortion of the atomic crystal structure of the GeV center, which would alter the energy levels and thus the peak position[28]. We evaluated eleven GeV single photon emitters in this study. Four of them possessed peak positions largely shifted from 602 nm (see Supplementary Fig. S4), which should be caused by the implantation damages. The difference in the count rate in Fig. 3b occurs probably for the same reason. Two approaches can be considered to overcome the problem. (1) In the same manner as the ensemble, the MPCVD fabrication would provide GeV single photon sources with a uniform peak position and a narrower line width, potentially achieving indistinguishability as demonstrated for the SiV centers[28]. (2) A higher temperature annealing after ion implantation can reduce the damages. Our preliminary studies show it has a positive effect on the GeV center ensemble, but also suggest that further optimization is required (see Supplementary Section A).

From the viewpoint of the crystal structure and measured luminescent characteristics, the GeV and SiV centers have the similar features as a photon source. It is, however, expected that the GeV center has an advantage of more controlled fabrication by MPCVD. Even though the Ge solid source was placed with the diamond substrate on the sample holder and a Si source was not introduced intentionally during the MPCVD growth in this study, the concentration of Si in the diamond film was higher than that of Ge (see Supplementary Fig. S3). Silicon is a common contaminant due to its presence in silica parts such as windows. It is also possible that the larger atomic size of the Ge atom might reduce the incorporation efficiency into diamond compared with the Si atoms. This fact suggests that the GeV centers have a potential to be fabricated with high controllability for the scalable single photon sources, and could be less affected by the unintentional doping.

In summary, we have discovered a novel GeV color center in diamond and demonstrated it as a single photon emitter at room temperature with a ZPL at around 602 nm and an estimate of an excited-state lifetime of about 1.4 ns. The ion implantation technique and subsequent high-temperature anneal formed both ensemble and single photon emitters of the GeV centers, while diamond growth by MPCVD with a Ge crystal enabled us to fabricate high-quality GeV center ensemble. The first-principles calculation predicted that the GeV center has the same split-vacancy crystal structure as the SiV center, but shows the emission with the shorter wavelength resulting from the higher $e_g$ state in the energy level in the GeV center.

## Methods

### Sample preparation

High purity IIa-type (001) single-crystal diamond substrates (Element six, electronic grade) were used for the ion implantation experiments. The amount of nitrogen impurities in the substrate is below 5 ppb. Implantation of the Ge ions was performed by an ion implantation system over whole the surface at room temperature. The ion implantation energy ranged from 150 to 260 keV, and the Ge ion doses were $3.5\times10^8 - 5.9\times10^{13}$ cm$^{-2}$. The Ge ions ($^{70}$Ge or $^{73}$Ge) were implanted through mass separation. Subsequently, the samples were annealed at 800 ºC for 30 min. The CVD growth was performed by a MPCVD system with a spherical-form resonator. A gas mixture of H$_2$ (198 sccm) and CH$_4$ (2 sccm) was used for the diamond film growth, and a Ge crystal was placed together with a (111) diamond substrate (Sumitomo, Ib-type) on a Mo sample holder. The gas pressure and growth temperature were 6 kPa and 820 ºC, respectively. The growth was performed for 2 h, leading to a diamond film thickness of about 100 nm.

**Optical measurement**

The PL spectra and intensity mapping at room temperature were recorded by a micro-Raman system or a home-built confocal microscope set-up with an excitation wavelength of 532 nm. For the low temperature measurements at 10 K, a micro-PL system with an excitation wavelength of 532 nm was used. The $g^2(\tau)$ function was measured using a Hanbury Brown-Twiss interferometer[29] with two avalanche photo diode detectors. For the measurements of the confocal images and antibunching of the single centers, an oil-immersion objective (NA= 1.4) was used and a band pass filter (Edmund; 25 nm bandpass 600 nm) was used to avoid the Raman signals around ZPL.

**Theoretical calculation**

For performing the first-principles calculations, we employed DFT[30,31] with plane wave basis set. The exchange-correlation energy was expressed by the generalized gradient approximation (GGA) with PBE functional[32]. A 216-atom cell was used for the super-cell calculation. Four irreducible k-points including high-symmetric $\Gamma$, X, and L-points were used for momentum-space integration. Interaction of valence wavefunction and ions was expressed by Troullier-Martins type pseudopotentials[33].

**Acknowledgments**

This work was supported in part by The Kurata Memorial and Hitachi Science and Technology Foundation and by JSPS KAKENHI 24650278.

**References**

1. Gisin, N., Ribordy, G., Tittel, W. & Zbinden, H. Quantum cryptography. *Rev. Mod. Phys.* **74,** 145 (2002).

2. Bouwmeester, D. et al., A. Experimental quantum teleportation. *Nature* **390,** 575 (1997).

3. Zaitsev, A. M. Vibronic spectra of impurity-related optical centers in diamond. *Phys. Rev. B* **61,** 12909 (2000).

4. Aharonovich, I. et al., Diamond-based single-photon emitters. *Rep. Prog. Phys.* **74,** 076501 (2011).

5. Pezzagna, S., Rogalla, D., Wildanger, D., Meijer, J. & Zaitsev, A. Creation and nature of optical centres in diamond for single-photon emission-overview and critical remarks. *New J. Phys.* **13,** 035024 (2011).

6. Gruber, A. et al., Scanning confocal optical microscopy and magnetic resonance on single defect centers. *Science* **276,** 2012 (1997).

7. Brouri, R., Beveratos, A., Poizat, J-P. & Grangier, P. Photon antibunching in the fluorescence of individual color centers in diamond. *Opt. Lett.* **25,** 1294 (2000).

8. Mizuochi, N. et al., Electrically driven single photon source at room temperature in diamond. *Nature Photon.* **6,** 299 (2012).

9. Wang, C., Kurtsiefar, C., Weinfurter, H. & Burchard, B. Single photon emission from SiV centres in diamond produced by ion implantation. *J. Phys. B: At. Mol. Opt. Phys.* **39,** 37 (2006).

10. Neu, E. et al., Single photon emission from silicon-vacancy colour centres in chemical vapour deposition nano-diamonds on iridium. *New J. Phys.* **13**, 025012 (2011).

11. Gaebel, T. et al., Stable single-photon source in the near infrared. *New J. Phys.* **6,** 98 (2004).

12. Aharonovich, I. et al., Two-level ultrabright single photon emission from diamond nanocrystal. *Nano Lett.* **9,** 3191 (2009).

13. Prawer. S. & Aharonovich, I. *Quantum Information Processing with Diamond: Principles and Applications*. ch. 6 (Woodhead Publishing, 2014).

14. Ziegler, J. P., Biersack, J. & Ziegler, M. D. SRIM – The Stopping and Range of Ions in Matter. (Lulu Press Co., 2008).

15. Leifgen, M. et al., Evaluation of nitrogen- and silicon-vacancy defect centres as single photon sources in quantum key distribution. *New J. Phys.* **16,** 023021 (2014).

16. Babinec, T. M. et al., A diamond nanowire single-photon source. *Nature Nanotechnol.* **5,** 195 (2010).

17. Rogers, L. J. et al., Multiple intrinsically identical single-photon emitters in the solid state.

*Nature. Comm.* **5,** 4739 (2014).

18. Magyar, A. et al., I. Syntheis of luminescent europium defects in diamond. *Nature. Comm* **5,** 3523 (2014).

19. Goss, J. P., Briddon, P. R., Rayson, M. J., Sque, S. J. & Jones, R. Vacancy-impurity complexes and limitations for implantation doping of diamond. *Phys. Rev. B* **72,** 035214 (2005).

20. Goss, J. P., Jones, R., Breuer, S. J., Briddon, P. R. & Oeberg, S. The Twelve-Line 1.682 eV Luminescence Center in Diamond and the Vacancy-Silicon Complex. *Phys. Rev. Lett.* **77,** 3041 (1996).

21. Clark, C. D., Kanda, H., Kiflawi, I. & Sittas, G. Silicon defect in diamond. *Phys. Rev. B* **51,** 16681 (1995).

22. Hepp, C. et al., Electronic structure of the silicon vacancy color center in diamond. *Phys. Rev. Lett.* **112,** 036405 (2014).

23. Rogers, L. J. et al., Electronic structure of the negatively charged silicon-vacancy center in diamond. *Phys. Rev. B* **89,** 235101 (2014).

24. Gali, A. & Maze, J. R. *Ab initio* study of the split silicon-vacancy defect in diamond: Electronic structure and related properties. *Phys. Rev. B* **88,** 235205 (2013).

25. Despite ignorance of the spin polarization in the present calculation and difference in the cell size (256 atoms in the present work and 512 atoms in Ref. 24) as well as k-point sampling (four k-points at the wedge of the whold Brillouin zone in the present work and Γ point in Ref. 24), the agreement in the $e_u$ and $e_g$ energy levels between the present calculation and Ref. 24 are satisfactory good. We thus judge that quantitative results provided by the present computational scheme is enough to discuss impurity levels.

26. Faklaris, O. et al., Detection of single photoluminescent diamond nanoparticles in cells and study of the internalization pathway. *small* **4**, 2236 (2008).

27. Igarashi, R. et al., Real-time background-free selective imaging of fluorescent nanodiamonds in vivo. *Nano Lett.* **12**, 5726 (2012).

28. Sipahigil, A. et al., Indistinguishable photons from separated silicon-vacancy centers in

diamond. *Phys. Rev. Lett.* **113,** 113602 (2014).

29. Hanbury Brown R. & Twiss, R. Q. A test of a new type of stellar interferometer on Sirius. *Nature* **178,** 1046 (1956)

30. Hohenberg, P. & Kohn, W. Inhomogeneous electron gas. *Phys. Rev.* **136,** 864 (1964).

31. Kohn, W & Sham, L. Self-consistent equations including exchange and correlation effects. *Phys. Rev.* **140,** 1133 (1965).

32. Perdew, J. P., Burke, K. & Ernzerhof, M. Generalized gradient approximation made simple. *Phys. Rev. Lett.* **77,** 3865 (1996).

33. Troullier, N. & Martins, J. L. Efficient pseudopotentials for plane-wave calculations. *Phys. Rev. B* **43,** 1993 (1991).

# Supplementary Information

# A Germanium-Vacancy Single Photon Source in Diamond

Takayuki Iwasaki,[1,2] Fumitaka Ishibashi,[3] Yoshiyuki Miyamoto,[2,4] Yuki Doi,[3] Satoshi Kobayashi,[3] Takehide Miyazaki,[2,4] Kosuke Tahara,[1] Kay D. Jahnke,[5] Lachlan J. Rogers,[5] Boris Naydenov,[5] Fedor Jelezko,[5] Satoshi Yamasaki,[2,6] Shinji Nagamachi,[7,8] Toshiro Inubushi,[8] Norikazu Mizuochi[2,3] and Mutsuko Hatano[1,2]

[1]Department of Physical Electronics, Tokyo Institute of Technology, Meguro, Tokyo 152-8552, Japan

[2]CREST, Japan Science and Technology Agency, Chiyoda, Tokyo

[3]Graduate School of Engineering Science, Osaka University, Toyonaka, Osaka 560-8531, Japan

[4]Nanosystem Research Institute, National Institute of Advanced Industrial Science and Technology, Tsukuba, Ibaraki 305-8568, Japan

[5]Institute for Quantum Optics and Center for Integrated Quantum Science and Technology (IQst), Ulm University, Albert-Einstein-Allee 11, Ulm D-89081, Germany

[6]Energy Technology Research Institute, National Institute of Advanced Industrial Science and Technology, Tsukuba, Ibaraki 305-8568, Japan

[7]Nagamachi Science Laboratory, Amagasaki, Hyogo 661-0976, Japan

[8]Shiga University of Medical Science, Otsu, Shiga 520-2192, Japan

**A. Effect of the Ge dose on the PL intensity.**

The dependence of the emission intensity from the GeV centers on the Ge dose during ion implantation was investigated. Figure S1 shows PL spectra with the Ge doses ranging from $2\times10^{12}$ – $2\times10^{15}$ cm$^{-2}$, prepared in a polycrystalline diamond substrate (Element six) by a focused-ion beam system with an ion energy of 40 keV. Apparently, the higher Ge doses lead to the higher ZPL intensities at 602 nm. Annealing at 1000 ºC after ion implantation reduced the line width and histogram of the peak position compared to annealing at 800 ºC (shown in Fig. 1 and 4 in the main text). However, the surface of the diamond substrates was sometimes graphitized in part at this higher temperature, resulting in reduced and non-uniform fluorescence intensities. Thus, we note that the optimization of the annealing conditions is necessary in further study.

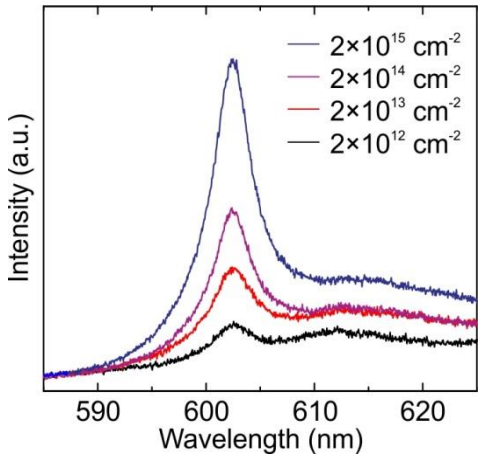

**Figure S1.** Dependence of the PL spectra on the Ge dose.

**B. Effect of high-temperature annealing on the formation of the GeV centers.**

The high-temperature anneal has to be performed to fabricate the GeV centers by ion implantation. Ge ion implantation on its own does not produce the characteristic GeV ZPL, while the spectrum after annealing at 800 ºC shows a sharp ZPL.

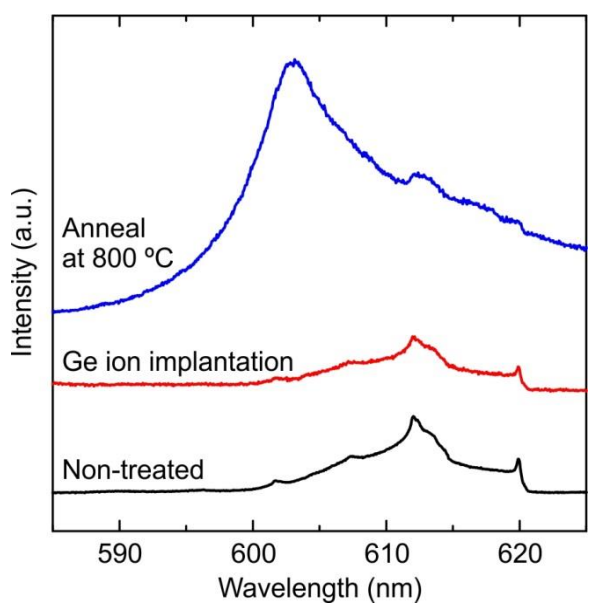

**Figure S2.** Dependence of PL spectra on the preparation process: non-treated single-crystal diamond substrate, after the Ge implantation, and after annealing at 800 ºC.

## C. SIMS profile of CVD-grown diamond film containing GeV centers.

Figure S3 shows secondary ion mass spectrometry (SIMS) profiles from the diamond film containing the GeV centers incorporated by MPCVD. Since a Ib-type substrate contains a large amount of N atoms, we can distinguish the CVD-grown film from the substrate. From the surface (depth=0 nm) to a depth of around 100 nm, the N concentration is under the detection limit. This region corresponds to the CVD film synthesized with a growth rate of 50 nm/h. At the interface at 100 nm, the concentrations of Ge and Si atoms become high due to easier incorporation of impurities at the interface, which is frequently observed for various dopants in diamond[1,2]. Although the Ge concentration fluctuates with depth, the film clearly contains the Ge atoms with a concentration of $1 - 6\times10^{16}$ cm$^{-3}$. In contrast, a higher concentration of Si atoms ($\sim10^{18}$ cm$^{-3}$) is incorporated in the film from contaminations and/or quartz parts in the growth chamber.

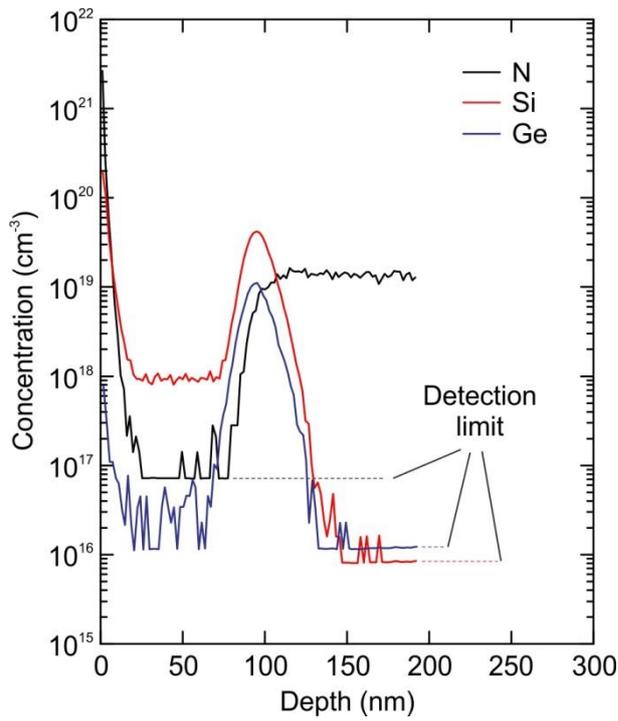

**Figure S3.** SIMS profiles from a diamond film containing GeV centers.

### D. ZPL peak position and intensity of GeV single emitters.

Eleven single GeV emitters fabricated by ion implantation were measured in this study. We found the variation of the ZPL peak position and ZPL peak intensity in the PL spectra (Fig. S4), which is probably caused by the damages created during ion implantation.

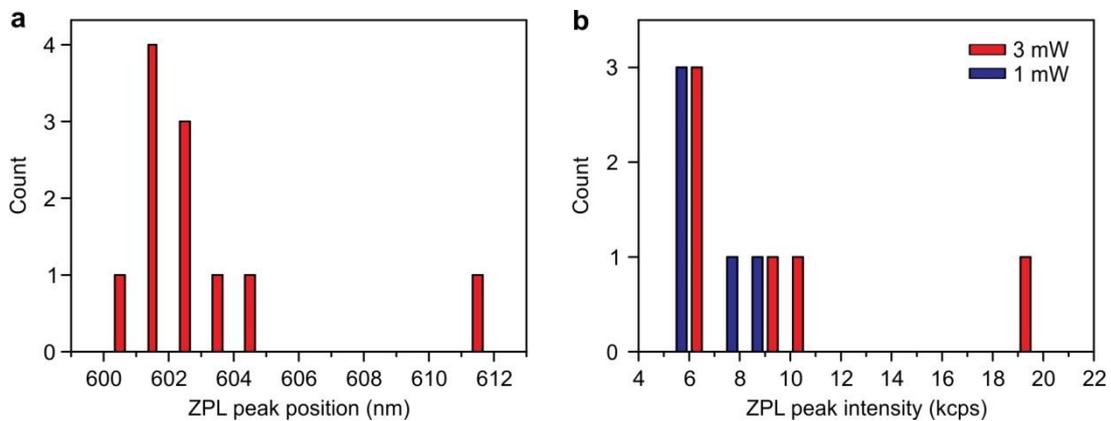

**Figure S4.** Variation of (a) ZPL peak position and (b) ZPL peak intensity of the GeV single emitters. The data were obtained from PL spectra. The peak intensities at the different laser powers are shown separately.

**References**

1. Makino, T., Kato., Ri, S-G., Yamasaki, S. & Okushi, H. Homoepitaxial diamond $p$-$n^+$ junction with low specific on-resistance and ideal built-in potential. *Diamond Relat. Mater.* **17,** 782 (2008).

2. Makino, T., Kato, H., Tokuda, N., Ogura, M., Takeuchi, D., Oyama, K., Tanimoto, S., Okushi, H. & Yamasaki, S. Diamond Schottky-pn diode without trade-off relationship between on-resistance and blocking voltage. *Phys. Status Solidi A* **207,** 2105 (2010).